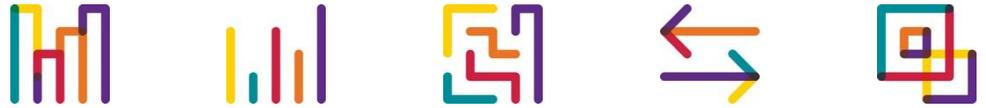

**WHITE PAPER ISSUE 5**

# Toward the "New Normal" – A Surge in Speeding, New Volume Patterns, and Recent Trends in Taxis/For-Hire Vehicles

Jingqin Gao, Abhinav Bhattacharyya, Ding Wang, Nick Hudanich, Siva Sooryaa
Muruga Thambiran, Suzana Duran Bernardes, Chaekuk Na, Fan Zuo, Zilin Bian
Kaan Ozbay, Shri Iyer, Hani Nassif, Joseph Y.J. Chow

Contact: c2smart@nyu.edu
c2smart.engineering.nyu.edu

## Executive Summary

Six months into the pandemic and one month after the phase four reopening in New York City (NYC), restrictions are lifting, businesses and schools are reopening, but global infections are still rising. This white paper updates travel trends observed in the aftermath of the COVID-19 outbreak in NYC and highlights some findings toward the "new normal."

## Key Findings

- According to MTA data (1), vehicular traffic volumes at major NYC crossings have bounced back to only 10% less than pre-pandemic levels by the week of August 17. Previous editions of this white paper used a MATSim simulation framework to estimate the return of traffic and travel with NYC's phased reopening. Estimated results compared with observed VMT data show that simulation predicted the VMT restoration in Phase 4 within 6% of observed VMT.
- The rebound in VMT is uneven across NYC, with some boroughs seeing a slow comeback, and some boroughs experiencing a higher VMT than the pre-pandemic period. In Queens, VMTs were observed to be 18% lower in August than February, while VMT in Manhattan, Brooklyn, Bronx and Staten Island in August were 27%, 51%, 32%, and 23%, higher than February, respectively.
- Estimated bus ridership and scheduled Access-A-Ride trips were down 35% and 36%, while subway and commuter rail ridership continued to lag, down 75% from 2019.
- Travel times on the 495 Corridor (2) slightly increased by an average of 6% eastbound and 4% westbound in the first week of August, compared with the first week of July. However, these travel times are still about 20% lower in both directions compared to pre-pandemic levels.
- In April 2020, the total number of monthly rides for yellow taxis dropped 96%, 92% for green taxis, 79% for for-hire vehicles (FHVs), and 76% for high volume for-hire services (HVFHS) as compared to January 2020, respectively. The number of rides started to increase gradually in May and June but are still 91%, 86%, and 63% lower for yellow, green, and HVFHS than in January 2020. The average distance of a completed ride increased from February to May, from 2.4 miles to 4.8 miles for yellow taxis, and 2.7 miles to 3.8 miles for green taxis. This, coupled with the drop in ridership, indicates that people are taking longer, but less frequent, trips.
- The average number of daily school zone speeding tickets issued remains 70% higher in July and 62% higher in the first two weeks of August 2020 compared with pre-pandemic levels in 2020. Speeding violations are also likely to occur repetitively: 23% of ticketed vehicles received more than one speeding ticket in June, and these vehicles contributed to 44% of total speeding tickets.
- Unsafe speed was listed as the primary cause for half of the traffic fatalities in NYC in April (7 out of 14) and 42% of the fatalities in May (5 out of 12) according to NYPD's Motor Vehicle Collision reports (12).
- Weigh-in-motion (WIM) data from C2SMART's testbed on the Brooklyn-Queens Expressway (BQE) (3) showed that average daily traffic (ADT) has fully recovery for both Queensbound (QB) and Staten Island-bound (SIB) traffic by the first two weeks of August, compared with February 2020. Average daily truck traffic (ADTT) for QB is 5% higher and is down by 8% SIB in August vs. February.
- The demand for cycling continued to increase in July. The daily number of Citi Bike trips (5) was 9% higher compared to June, and up 71% from February.
- Detection results based on a deep-learning based video-processing algorithm (4) show that average pedestrian density from 11 select locations in NYC continued to see a gradual increase (+10%) in July compared to June. The average percentage of pedestrians observed to be able to follow social distancing guidelines of remaining a minimum of 6 feet from others remained at a constant level (~82%) in June and July, as seen from traffic cameras.
- 24-hour density distribution estimations show that car and pedestrian densities over the course of the day resemble pre-pandemic patterns. The cyclist density in observed locations has exceeded pre-pandemic levels, for example up 34% in the afternoon (2-6:00 PM) at 5th Ave/42 St.
- Continued tracking of subway ridership from multiple cities in China show that only 85% of pre-pandemic ridership levels have been restored 6-months after full reopening of those cities. Where a second wave occurred, the restoration only reached 65% of pre-pandemic volume.

**Mobility Trends**

Vehicular Traffic and Transit Trends

The average daily vehicular traffic via MTA bridges and tunnels (B&T) continued to rebound through mid-August, bouncing back from pandemic lows to 90% of pre-pandemic levels, according to MTA data (1). Table 1 presents the weekly percentage change in vehicular traffic via MTA bridges and tunnels, compared to the same week in 2019. While most of the bridges and tunnels in Queens and Manhattan had a relatively slower recovery (e.g., -33% Queens Midtown Tunnel, -27% Henry Hudson Bridge, -20% Robert F. Kennedy Bridge in the week of August 9 compared to the same week in 2019), bridges and tunnels connecting Brooklyn were down only about 10%. Traffic volume remained mostly stable for most bridges and tunnels over the past four weeks (July 19 to August 15).

**Table 1** % Change of weekly vehicular traffic via MTA bridges and tunnels by plaza, 2020 vs. 2019, Source: MTA (1)

| Plaza | 7/12 | 7/19 | 7/26 | 8/2 | 8/9 |
|---|---|---|---|---|---|
| Bronx-Whitestone Bridge | -27% | -20% | -19% | -16% | -17% |
| Cross Bay Veterans Memorial Bridge | -15% | -5% | -7% | -8% | -11% |
| Henry Hudson Bridge | -35% | -28% | -28% | -27% | -27% |
| Hugh L. Carey Tunnel | -22% | -13% | -17% | -14% | -11% |
| Marine Parkway-Gil Hodges Mem. Bridge | -19% | -6% | -6% | -8% | -9% |
| Queens Midtown Tunnel | -40% | -35% | -36% | -34% | -33% |
| Robert F. Kennedy Bridge (Manhattan) | -28% | -23% | -24% | -22% | -21% |
| Robert F. Kennedy Brdg. (Queens/Bronx) | -26% | -18% | -21% | -19% | -20% |
| Throgs Neck Bridge | -17% | -16% | -19% | -17% | -23% |
| Verrazano-Narrows Bridge | -18% | -13% | -15% | -13% | -15% |

Bus ridership and scheduled Access-A-Ride trip volumes (1) were 35% and 36% of 2019 averages by the week of August 17, 2020, which were 20 and 17 percentage points increase from June, respectively. However, average subway and commuter rail ridership in NYC continues to stay low – a 74%, 73% and 77% decline in weekly ridership for NYC subway, Long Island Rail Road (LIRR), and Metro North Railroad (MNR) in the week of August 17, compared to weekly averages in 2019. Figure 1 shows the weekly ridership trends reported by MTA compared to 2019.

Travel times on the 495 Corridor (2) increased by an average of 6% eastbound and 4% westbound in the first week of August, compared with the first week of July 2020. However, these travel times are still about 20% lower in both directions compared to pre-pandemic levels.

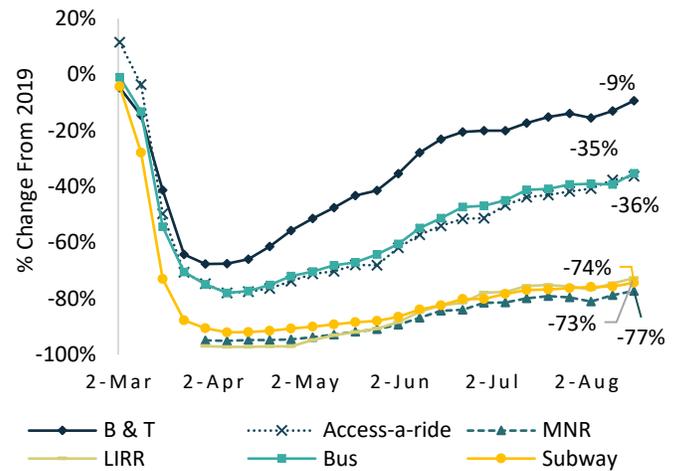

**Figure 1** Mobility trends in NYC compared to 2019 Weekday/Saturday/Sunday average, Source: MTA (1)

As of August 16, Weigh-in-motion (WIM) data from C2SMART's testbed on the Brooklyn-Queens Expressway (BQE) (3) showed that average daily traffic (ADT) is down by only 2% for Queensbound (QB), and 4% for Staten Island-bound (SIB) traffic in the first two weeks of August, as compared with February 2020. Average daily truck traffic (ADTT) for SIB is down by 8% and is 5% higher QB in August compared to February. The average vehicle speed on BQE remained around the same level in the first weeks of August compared to June but is still 8% higher for SIB and 2% higher for QB, as compared to February.

Pedestrians and Cycling

C2SMART researchers continued to investigate crowd density trends and social distancing patterns during the subsequent COVID-19 recovery process using a deep learning-based computer vision method introduced in the [previous issues](#) of this white paper. Object detection is being applied to real-time traffic camera videos at multiple key locations within NYC (4). In July, average pedestrian density continued to see a gradual increase (+10%) from 11 locations in NYC on selective days as compared with June (a 62% increase from April lows). The average percentage of pedestrians following social distancing guidelines of remaining a minimum of 6 feet from others remained the same (~82%) in June and July.

Figure 2 illustrates the 24-hour temporal density of daily snapshots of the distributions of pedestrian, cars, and cyclists at a selected location (5 Ave/42 St, Manhattan). A gradual increase was observed in all three modes. The temporal distribution in car and pedestrian densities around the end of July were observed to be similar to pre-pandemic patterns over the course of the day, whereas the cyclist density exceeded pre-pandemic levels with a 34% uptick in the afternoon period (2:00 PM– 6:00 PM) at this location.

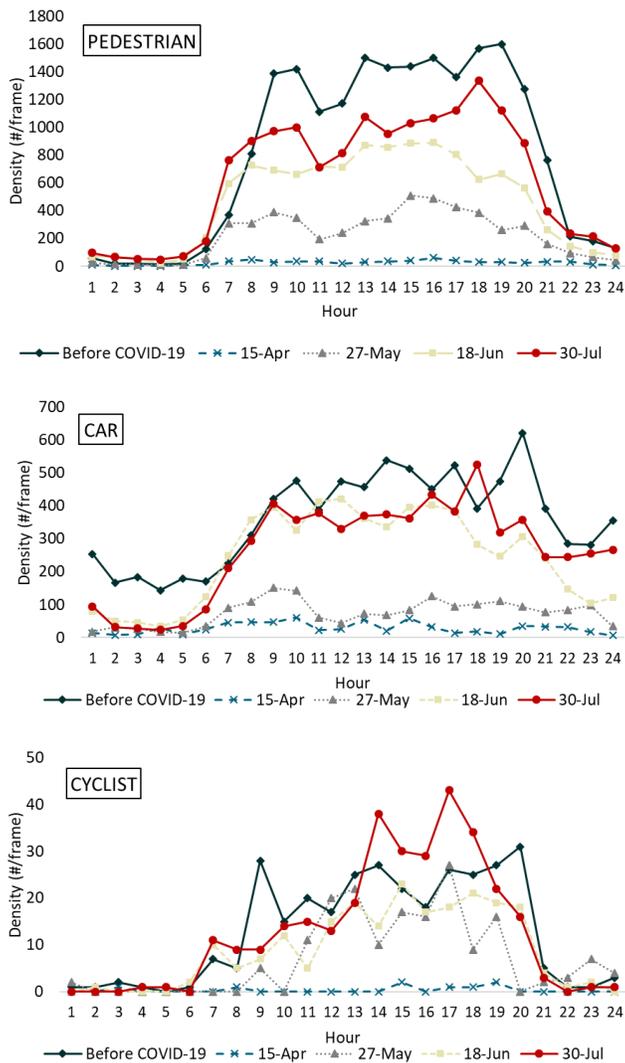

**Figure 2** Temporal distributions of pedestrian, car, and cyclist densities (5 Ave/42 St, Manhattan)

Cycling continued to increase in July. According to Citi Bike system data (5), on average, there were 67,422 rides in NYC per day in July, a 9% increase compared to June and a 71% increase compared to February 2020. Figure 3 illustrates the Citi Bike daily ridership trends since March.

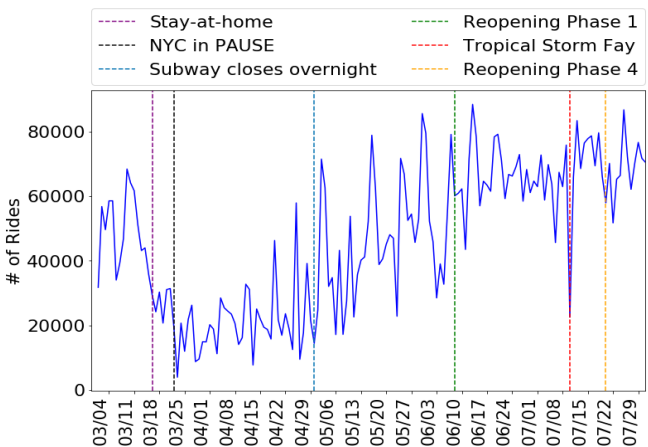

**Figure 3** Citi Bike Ridership Trends, Source: Citi Bike (5)

## Uneven Increase in Vehicle Miles Travelled

According to Streetlight (7), total Vehicle Miles Travelled (VMT) in the NYC region has been steadily increasing since the initial phases of reopening. Nevertheless, the VMT increase is uneven across NYC, with some boroughs seeing a slow comeback, and some boroughs experiencing even a higher VMT than the pre-pandemic period (Figure 4). In Queens, VMTs were observed to be 18% lower in August than February 2020. However, VMT in other boroughs in August surpassed pre-pandemic levels.

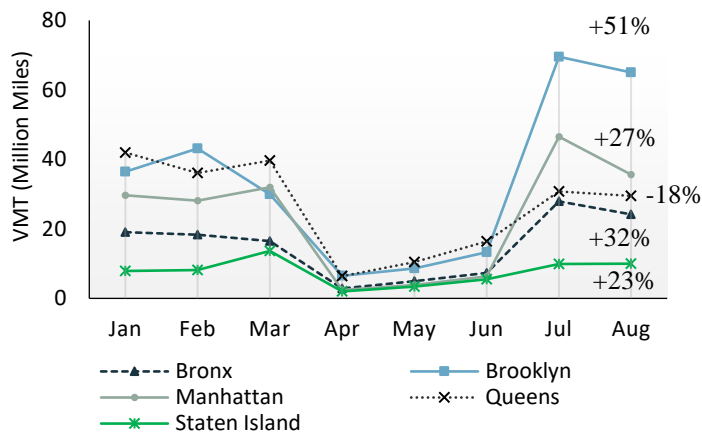

**Figure 4** Monthly NYC VMT trends by borough (estimated from the 1st week of each month), August vs. February 2020, Source: StreetLight (7)

### Interborough Traffic

StreetLight InSight (7) data was used to estimate the change in interborough traffic in April 2020 and July 2020 compared with April 2019 levels in Table 2 and Table 3. By July, interborough traffic has seen a nearly full recovery from April lows, except trips originating and ending in Manhattan, which saw a much slower rebound. Moreover, traffic between Brooklyn and the Bronx has increased (41% and 43%) from pre-pandemic levels.

**Table 2** Comparison of borough-to-borough total traffic, Source: StreetLight (7)

| All vehicles – April 2020 vs April 2019 | | | | | |
|---|---|---|---|---|---|
| Orig./Dest. | Bronx | Bklyn. | Manh. | Queens | Staten Is. |
| Bronx | -45% | -22% | -44% | -44% | -28% |
| Brooklyn | -22% | -54% | -65% | -51% | -55% |
| Manhattan | -42% | -62% | -72% | -64% | -60% |
| Queens | -39% | -46% | -62% | -53% | -43% |
| Staten Is. | -24% | -52% | -61% | -46% | -54% |
| All vehicles – July 2020 vs April 2019 (*July 2020 data is based on the first two weeks) | | | | | |
| Orig./Dest. | Bronx | Bklyn. | Manh. | Queens | Staten Is. |
| Bronx | -6% | 43% | -3% | -9% | 3% |
| Brooklyn | 41% | 3% | -23% | -3% | -14% |
| Manhattan | -3% | -22% | -49% | -32% | -43% |
| Queens | -2% | 3% | -29% | -17% | -12% |
| Staten Is. | 7% | -7% | -36% | -11% | -35% |

For freight traffic, an earlier rebound was found for both inbound and outbound traffic in Staten Island in June, possibly due to the increased delivery demand from the Amazon Fulfillment Center. Currently freight data is only available up to June.

**Table 3** Comparison of borough-to-borough freight traffic, Source: StreetLight (7)

| Freight only – April 2020 vs April 2019 | | | | | |
|---|---|---|---|---|---|
| Orig./Dest. | Bronx | Bklyn. | Manh. | Queens | Staten Is. |
| Bronx | -65% | -27% | -64% | -39% | -8% |
| Brooklyn | -24% | -65% | -59% | -58% | -32% |
| Manhattan | -64% | -62% | -73% | -68% | -37% |
| Queens | -41% | -58% | -66% | -64% | -30% |
| Staten Is. | -3% | -30% | -37% | -33% | -56% |
| Freight only – June 2020 vs April 2019 | | | | | |
| Orig./Dest. | Bronx | Bklyn. | Manh. | Queens | Staten Is. |
| Bronx | -45% | 6% | -40% | -7% | 18% |
| Brooklyn | 9% | -37% | -28% | -23% | 22% |
| Manhattan | -41% | -35% | -54% | -38% | -6% |
| Queens | -8% | -24% | -35% | -43% | -4% |
| Staten Is. | 28% | 26% | 4% | -7% | -29% |

## Taxis and For Hire Vehicles

The recent data released by the NYC Taxi and Limousine Commission (TLC) (9) show the sharp decrease in yellow taxi, green taxi, FHV, and high volume for-hire services (HVFHS) usage from the second week of March 2020 onwards. FHVs include Community Cars (aka Liveries), Black Cars, and Luxury Limousines, whereas HVFHS include records from Uber, Lyft, and Via. At the beginning of April, trips in all TLC licensed vehicles dropped 84% compared to the same time last year. According to a report by TLC (8), in April 2020, 28,893 TLC drivers were on the road, compared with 122,076 TLC drivers in April 2019 (a 76% drop). From March through April 2020, trips to parts of the city with hospitals increased as a percentage of all pick-ups, while trips to and from airports decreased (8).

### Trip Volume

According to TLC data (9), the total number of monthly rides for yellow taxis dropped by 96%, 92% for green taxis, 79% for FHVs, and 76% for HVFHS in April compared to January 2020 (Table 4). The number of rides gradually rebounded in May and June but are still 60-90% lower than those in January 2020. In February, HVFHSs completed, on average, nearly 750,000 trips per day, Yellow Taxis completed 217,000 trips per day, and Green Taxis completed 14,000 trips per day. Trips began to decline for all three types in the middle of March and reached their lowest point in April (144,000, 8,000, and 1,000 trips per day, respectively) (Figure 5).

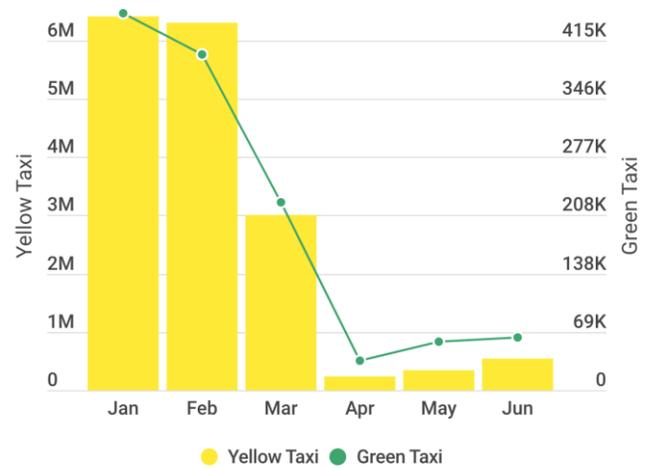

(a) Total number of rides per month from January to June 2020 for Yellow and Green Taxis

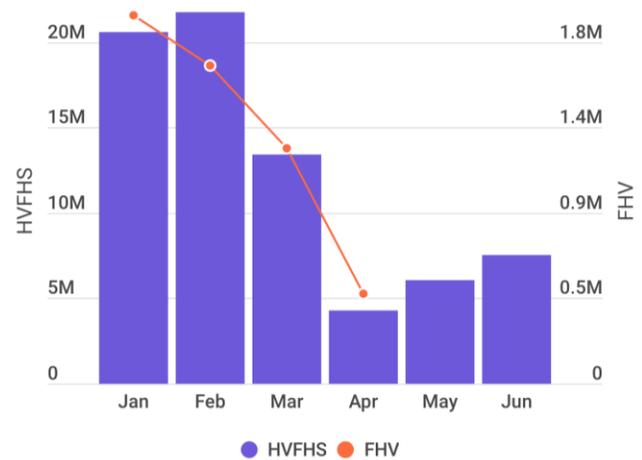

(b) Total number of rides per month from January to June 2020 for FHV and HVFHS

**Figure 5** Yellow/Green Taxi, FHV and HVFHS rides from January to June 2020, Source: TLC (9)

**Table 4** Percentage changes in monthly ridership levels using January 2020 as baseline, Source: TLC (9)

| Month | Yellow Taxi | Green Taxi | HVFHS | FHV |
|---|---|---|---|---|
| **Feb** | -2% | -11% | 6% | -14% |
| **Mar** | -53% | -50% | -35% | -36% |
| **Apr** | -96% | -92% | -79% | -76% |
| **May** | -95% | -87% | -70% | |
| **Jun** | -91% | -86% | -63% | |

### Spatial Distributions

Figure 6 shows the spatial distribution in yellow taxi, green taxi, and HVFHS rides from March to June 2020. Green taxi rides ending in the upper Manhattan zones of Central Harlem North, Hamilton Heights, and Washington Heights North increased in June 2020. HVFHS had seen the most recovery in ridership levels since the start of the pandemic at JFK airport, Forest Hills, and Middle Village, Queens. For the Bronx and Staten Island, total rides have remained steady in the last four months.

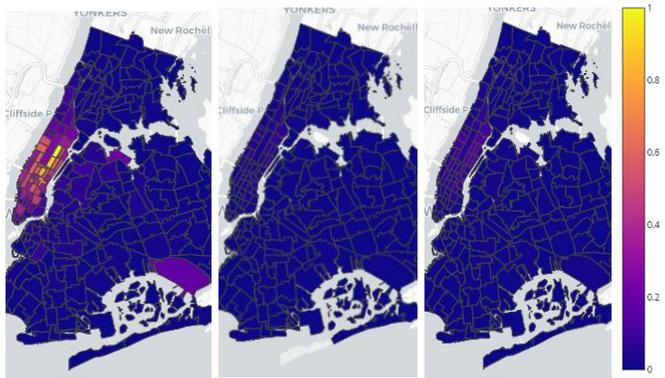

(a) Yellow Taxis (left-March, middle-April, right-June 2020)

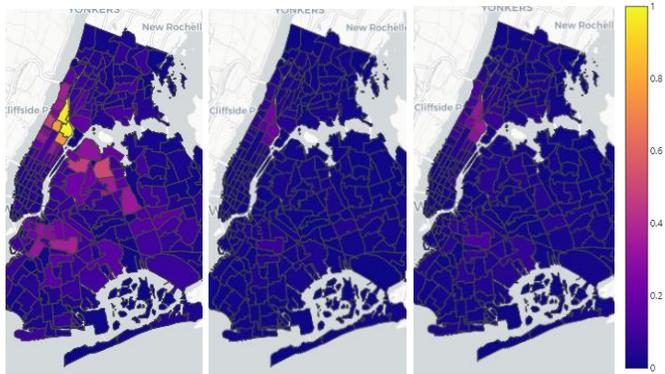

(b) Green Taxis (left-March 2020, right-June 2020)

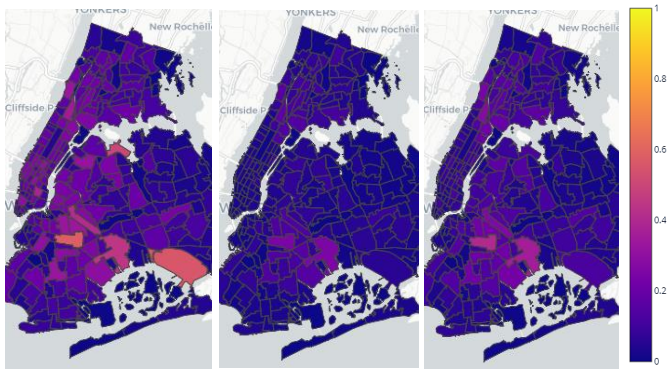

(c) HVFHS (left-March, middle-April, right-June 2020)

**Figure 6** Spatial distributions by sector

## Trip Distance, Duration and Passenger Count

The average distance of a completed rides has increased during the pandemic (Figure 7). This, coupled with the drop in ridership, points to a trend where people are taking longer trips, but perhaps only when most necessary. As ridership numbers increase, the average trip distance also become shorter.

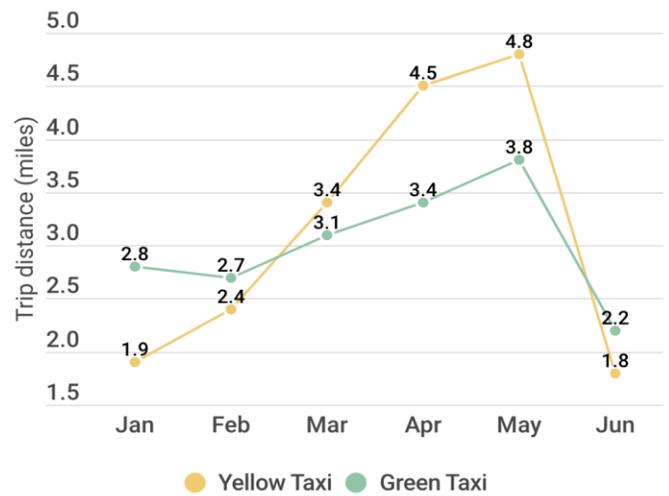

**Figure 7** Yellow/Green Taxi Trip Distance, Source: TLC (9)

The average duration of a trip increased slightly during the pandemic and is gradually seen to be nearing pre-pandemic levels by June 2020.

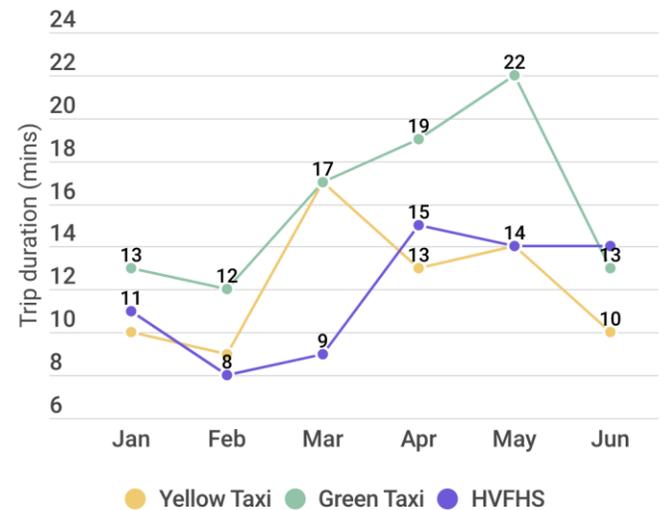

**Figure 8** Yellow/Green Taxi Trip Duration, Source: TLC (9)

The average number of passengers decreased, especially for green taxis, since the start of the pandemic - this might be because the passengers are advised against sitting beside the driver in any taxis or FHVs, to adhere to social distancing norms. The average passenger count for a ride for the months of January to June 2020 are shown in Figure 9.

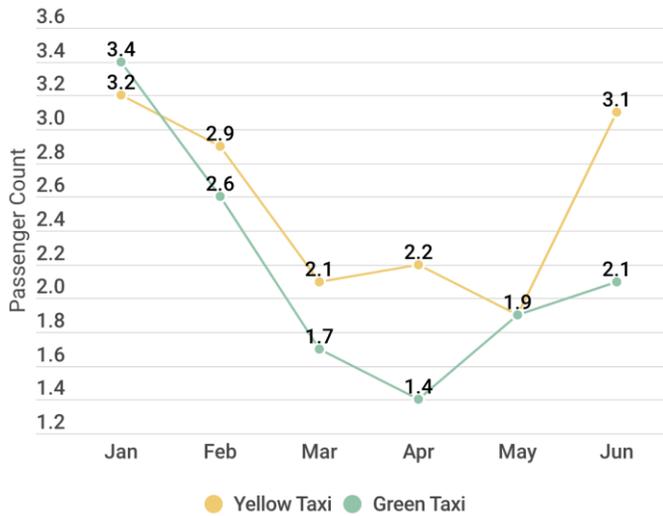

**Figure 9** Yellow/Green Taxi Passenger Counts, Source: TLC (9)

**A Surge in Speeding**

While the pandemic has resulted in reductions in traffic congestion, speeding, observed by the number of speeding tickets issued, has increased. Data from speed camera violations (10) highlights that speeding violations skyrocketed during the pandemic. The New York Independent Budget Office reported that although parking violations were issued despite restrictions being lifted during the March 23 to May 31 period in Question, 77% of all summonses issued were for speed (11). Average daily school zone speeding tickets were up 70% and 62% in July and the first two weeks of August 2020 compared to January 1 to March 12 of 2020. Figure 10 presents the time series of weekly numbers of speeding tickets since January 2020.

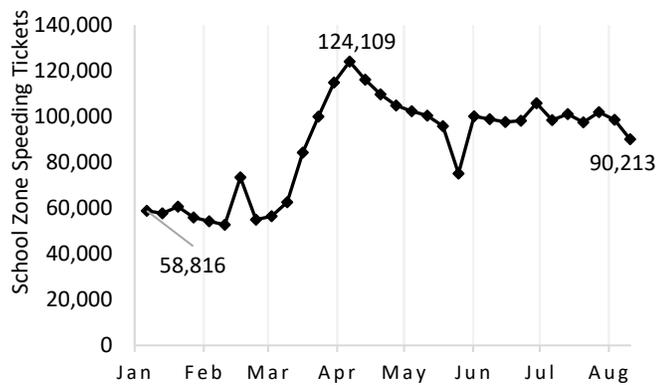

**Figure 10** School Zone Speeding Ticket Trend (weekly sum), Source: Open Parking and Camera Violations (10)

Figure 11 illustrates the temporal distributions of the speeding tickets from February to June. The findings indicate that afternoon (2:00 PM – 6:00 PM) is the period with the highest number of speeding violations.

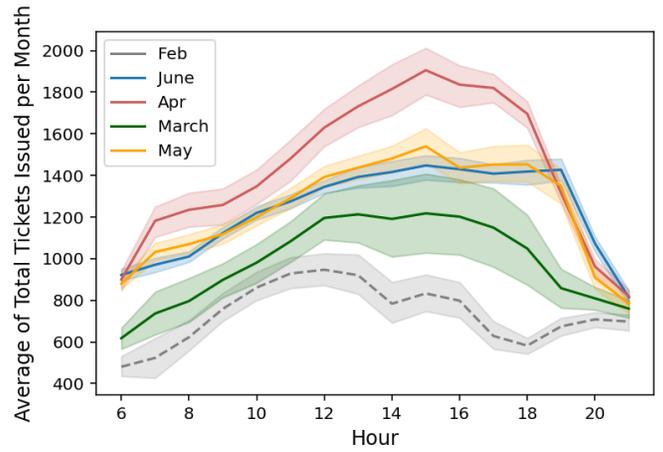

**Figure 11** Average of total tickets issued per month by hour, Source: Open Parking and Camera Violations (10)

Speeding Tickets by Vehicle Type

After examining unique plate IDs, speeding tickets per vehicle is summarized in Table 5. 14%, 29%, and 23% of the vehicles received more than one speeding ticket in February, April, and June 2020, respectively, and these vehicles contributed to 27%, 54%, 44% of the total speeding tickets in February, April, and June 2020, respectively.

**Table 5** Speeding tickets per vehicle, Source: Open Parking and Camera Violations (10)

|  | >1 tickets | >3 tickets | >5 tickets | >10 tickets |
|---|---|---|---|---|
| **Feb** | 13.6% | 0.8% | 0.1% | 0.0% |
| **Apr** | 29.1% | 5.6% | 1.7% | 0.2% |
| **Jun** | 22.8% | 3.2% | 0.8% | 0.1% |

Table 6 lists the top 10 plate types in February, April and June that had the highest number of speeding tickets. Despite the overall reduction in speeding tickets from April to June, the total number of motorcycle speeding tickets continued to increase. In addition, a noticeable increase in utility vehicle speeding tickets was also observed in June. The total number of speeding tickets received by utility vehicles in June (20,823 tickets) is about five times higher than February (3,496 tickets).

**Table 6** Speeding tickets by plate type, Source: Open Parking and Camera Violations (10)

| Plate Type | Feb | Apr | Jun |
|---|---|---|---|
| PAS (Passenger Vehicles) | 207,818 | 449,256 | 301,349 |
| OMT (Taxi) | 11,273 | 20,554 | 12,004 |
| COM (Commercial Vehicles) | 4,945 | 11,144 | 8,732 |
| SRF (Special (Passenger)) | 3,858 | 7,515 | 4,977 |
| OMS (Special Omnibus Rentals) | 2,990 | 6,790 | 5,104 |
| ORG (Organization) | 982 | 1,872 | 1,180 |
| MED (Medical Doctor) | 549 | 1,269 | 692 |
| PSD (Political Subdivision) | 513 | 1,374 | 726 |
| SPO (Sports Series (Passenger)) | 435 | 1,000 | 623 |
| MOT (Motorcycles) | 370 | 1,581 | 1,763 |

## Speeding vs. Crashes

NYPD motor vehicle collision data (12) reports traffic crashes in NYC and the contributing factor(s) associated with each crash. Reported contributing factors include behaviors such as passing or lane usage, improper or unsafe lane changing, and unsafe speed. Figure 11 shows the total crashes and the number of crashes primarily due to unsafe speed reported by NYPD. It is worth mentioning that not all crashes were associated with a specific contributing factor, therefore the actual number of crashes due to unsafe speed could be higher than what was reported.

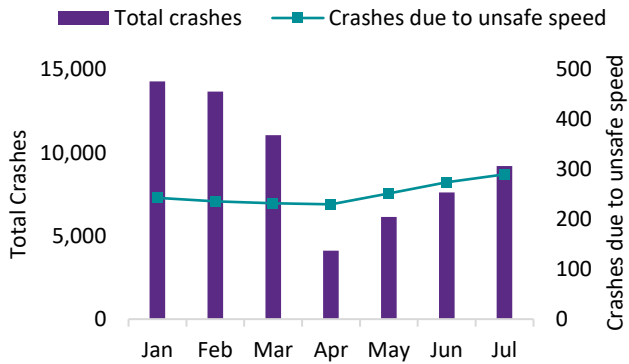

**Figure 12** Total crashes and crashes due to unsafe speed, Source: Motor Vehicle Collisions (12)

Despite a decrease in traffic-related fatalities due to lower traffic volumes and less exposure of pedestrians, the fatality rate in crashes (fatalities/number of total crashes) changed from 1.5 fatalities/1000 crashes in February 2020 to 3.4, 2.0, 4.2 and 2.5 fatalities/1000 crashes in April, May, June, and July 2020, respectively. Although a relative higher fatality rate was observed in April and June, more data is needed to identify a trend (Figure 13).

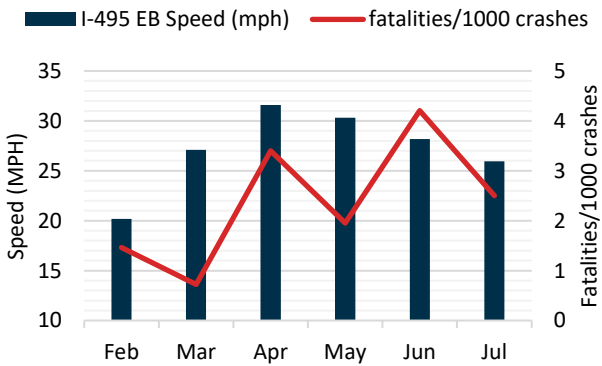

**Figure 13** Fatality rate vs speed, Source: Virtual sensors (2), Motor Vehicle Collisions (12)

Half of the fatalities in April (7 out of 14) and 42% of the fatalities in May (5 out of 12) are due to unsafe speed. June 2020 was the deadliest month for traffic deaths in nearly two years with 32 fatalities on New York City roadways, four of these being cyclists.

**Table 7** Fatalities for which unsafe speed is the primary cause, Source: Motor Vehicle Collisions (12)

|  | April | May | June | July |
|---|---|---|---|---|
| **Total fatalities** | 14 | 12 | 32 | 23 |
| **Fatalities due to unsafe speed** | 7 | 5 | 10 | 5 |
| **Cyclist fatalities** | 1 | 2 | 4 | 1 |

## Simulation Estimations

The June issue of this white paper introduced an agent-based simulation model (MATSim-NYC) that was used to predict how this pandemic could change travel behavior and provided some insights for the reopening of NYC. The VMT from MATSim simulation results is compared with the observed VMT data from StreetLight (7). The observed average daily VMT in July and August has been restored to 131% of the pre-pandemic period VMT (January 2020), while the predicted VMT restoration from the simulation model is 137% (a 6% difference). This shows that on an aggregate level the model was able to accurately predict the increased preference for driving by NYC's Phase 4 reopening.

Figure 14 shows the comparison of the predicted restored trips when NYC enters Phase 4 reopening with the most recently observed data. The daily transit ridership is based on MTA ridership (1) from July 20 to August 31, daily car and walk trip ratios are from the Apple mobility reports (13) from July 20 to August 30. CitiBike data is available from July 20 to August 2 from CitiBike system data (5). It should be noted that some industries which were planned to have fully reopened in phase 4 did not do so (e.g., university, indoor dining, etc.), therefore, predicted transit and car trips are slightly higher than the observed data. The difference in CitiBike trips is relatively large, but this might be due to several reasons, including the major CitiBike expansion this year as well as free membership for essential workers, which were not initially considered in the model assumptions.

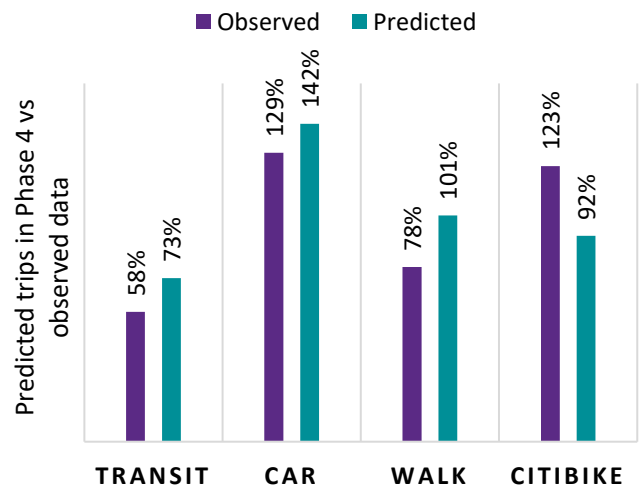

**Figure 14** Predicted restored trips in MATSim-NYC compared to observed data

Ongoing MATSim-NYC work will analyze the transportation system with different travel behavior and transit operation assumptions by examining the following performance measurements: traffic congestion, emissions, and social contacts in transit. 12 scenarios were designed for NYC traffic conditions in the post-pandemic period with considerations of four impact factors due to COVID-19: the change of mode preference, with/without transit capacity restriction, the rise of the remote workforce, and the flexible work hours suggested by the health authorities. Estimations will be reported in the future.

### Subway Ridership Recovery in China

To gain insights on what subway recovery might look like in the aftermath of COVID-19 and what may happen if a second wave occurs, subway ridership in multiple cities in China is shown in Figure 15. In Shanghai and Guangzhou, 85% of pre-pandemic ridership was restored 6 months after reopening. Wuhan, the Chinese epicenter of the COVID-19 outbreak, had a phased reopening one month after the other cities in China. It has seen a 61% recovery in subway ridership 5 months after reopening. Beijing implemented a stricter reopening strategy at first (e.g., mandate to maintain subway car occupancy below 50% of its maximum capacity) and also experienced a second wave of COVID-19 in late June. Beijing ridership has recovered to only 65% of pre-pandemic levels, 6 months after its initial reopening.

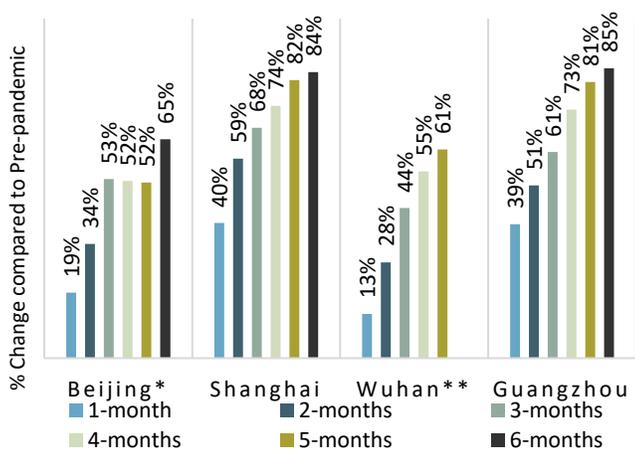

**Figure 15** Subway ridership changes in Guangzhou, Beijing, Shanghai and Wuhan, China (*Beijing had a second wave of COVID-19 in late June, ** Wuhan had a three-phased reopening one month after the reopening of other cities in China), Source: Official Metro Weibo (6)

### Summary of Findings

One month after the phase 4 reopening in NYC, traffic conditions are still evolving. The rebound in traffic is uneven across NYC and the recovery speed of different transportation subsystems is not uniform. New challenges such as the increasing VMT in certain regions, high demand for bikes, growing funding gaps for public transportation, and the surge in speeding behavior are raised. Adapting to the "new normal" and taking actions to address these enormous challenges in the aftermath of COVID-19 is crucial: providing a safe and accessible micromobility network and open space for pedestrians, reducing traffic violence and inequality, balancing vehicular traffic and encouraging sustainable transportation alternatives for all ages and abilities, and maintaining the affordability and functionality of transit systems. Deeper consideration for vulnerability assessments and incremental improvements to make transportation systems more resilient and well-prepared are still needed.

### Future Updates to COVID-19 Transportation Trends

C2SMART's COVID-19 Dashboard (available at http://c2smart.engineering.nyu.edu/covid-19-dashboard), an all-in-one mobility and sociability trend monitor, will continue to consolidate public data sources to track the impact of the pandemic on transportation systems. Most of the data and aggregated statistics will be updated weekly and are open for download. The current dashboard contains data from traditional traffic detectors, crowdsourcing applications, probe vehicles, real-time traffic cameras, and police and hotline reports from multiple affected cities. This platform will continue to evolve with the addition of new data, metrics, and visualizations. Through this interactive data dashboard, researchers, transportation authorities, and the general public have direct access to information that can be used to monitor the pandemic's ongoing impacts to our cities and transportation systems to make more effective data-driven decisions.

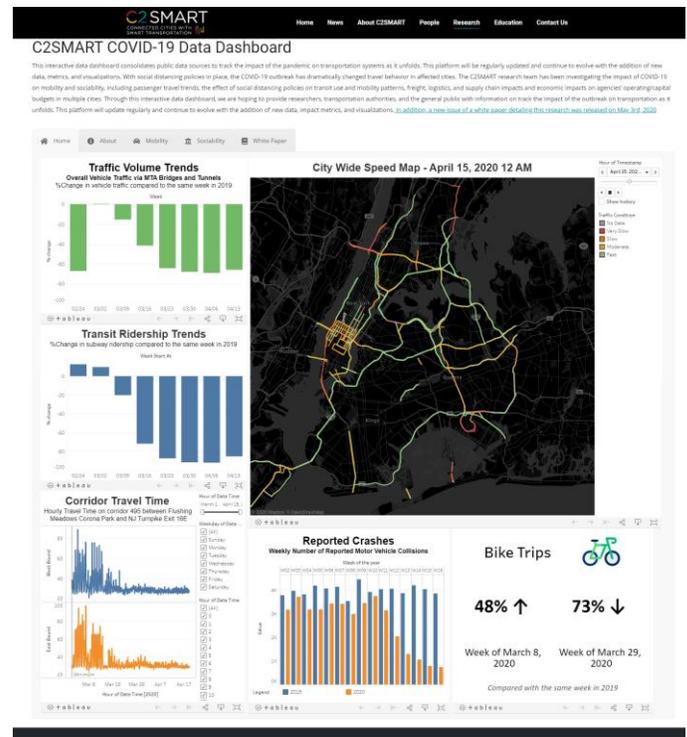

**Figure 16** C2SMART COVID-19 Data Dashboard

### References

1. MTA, Day-By-Day Ridership Numbers, https://new.mta.info/coronavirus/ridership
2. Morgul, Ender Faruk, et al. "Virtual sensors: Web-based real-time data collection methodology for transportation operation performance analysis." Transportation Research Record 2442.1 (2014): 106-116.

3. C2SMART Urban Roadway Testbed in Brooklyn NY, http://c2smart.engineering.nyu.edu/c2smart-roadway-urban-testbed/
4. Fan Zuo, Jingxing Wang, Jingqin Gao, Kaan Ozbay, Xuegang Jeff Ban, Yubin Shen, Hong Yang and Shri Iyer. (2020), An Interactive Data Visualization and Analytics Tool to Evaluate Mobility and Sociability Trends During COVID-19, UrbComp 2020: The 9th SIGKDD International Workshop on Urban Computing, San Diego, California, USA (accepted). http://urban.cs.wpi.edu/urbcomp2020/accept.html
5. Citi Bike System Data, https://www.citibikenyc.com/system-data
6. Shanghai, Beijing, Guangzhou and Wuhan Metro official Weibo account, Metro Ridership
7. StreetLight, VMT Monitor and StreetLight Insight, https://www.streetlightdata.com/
8. COVID-19 Impact on the NYC For-Hire Industry. Press release report by the Taxi and Limousine Commission, July 2020
9. The NYC Taxi and Limousine Commission Trip Record Data, https://www1.nyc.gov/site/tlc/about/tlc-trip-record-data.page
10. Open Parking and Camera Violations, NYC OpenData, https://data.cityofnewyork.us/City-Government/Open-Parking-and-Camera-Violations/nc67-uf89
11. QNS, Data shows speed camera violations have soared in New York City during lockdown, https://qns.com/story/2020/07/24/data-shows-speed-camera-violations-have-soared-in-new-york-city-during-lockdown/
12. Motor Vehicle Collisions – Crashes, NYC OpenData, https://data.cityofnewyork.us/Public-Safety/Motor-Vehicle-Collisions-Crashes/h9gi-nx95
13. Apple mobility trends report. https://covid19.apple.com/mobility

*Note: This paper reflects the Center's perspective as of September 3, 2020. All data is preliminary and subject to change. The authors gratefully acknowledge StreetLight for sharing its traffic data. For more information please contact c2smart@nyu.edu.*